\documentstyle[12pt]{article}
\newcommand{\tiN}{\raisebox{-6.5pt}{$\displaystyle
\stackrel{\displaystyle N}{\sim}$}}
\newcommand{\beq}{\begin{equation}}
\newcommand{\eeq}{\end{equation}}
\makeatletter
\@addtoreset{equation}{section}
\makeatother

\title{Spherically Symmetric Gravity as a Completely Integrable
System\thanks{Work supported by the Graduierten-Programm of the
DFG}}
\author{H.A.\ Kastrup\thanks{E-mail address: kastrup @
thphys.physik.rwth-aachen.de} and T.\ Thiemann\thanks{Present address: Center
for Gravitational Physics
and Geometry, Pennsylvania State University, University Park, PA 16802-6300,
USA;  E-mail address: thiemann @ phys.psu.edu}
     \\   Institute for Theoretical Physics, RWTH Aachen,\\
       52056 Aachen, FR Germany}
\date{{\small Preprint PITHA 93-35, November 93}}

\begin{document}
\maketitle
\begin{abstract} It is shown - in Ashtekar's canonical framework of General
Relativity - that  spherically symmetric (Schwarzschild) gravity in 4
dimensional space-time constitutes a
finite dimensional completely integrable system. Canonically conjugate
observables for asymptotically flat space-times are masses as action
variables and - surprisingly
- time variables as angle variables, each of which is associated with an
asymptotic "end" of the
Cauchy surfaces. The emergence of the time observable is a consequence of the
Hamiltonian formulation and its subtleties concerning the slicing of space and
time and is not in contradiction to Birkhoff's theorem. The results
 are of interest as to the concept of time in General
Relativity. They can be formulated within the ADM formalism, too.
Quantization of the system and the associated Schr\"odinger equation
depend on the allowed spectrum of the
masses. \end{abstract}
\section{Introduction} Recently we have shown\cite{1}(quoted as ref.\ I in
the following) that Ashtekar's constraints in his Hamiltonian formulation of
general relativity can be solved completely -
classically and quantum mechanically as well - for spherically symmetric field
configurations. As the two basic canonically conjugate observables - in the
sense of Dirac -  we
identified the mass squared $P=4m^2$ and a quantity called $Q$ which in
geometrical metric terms takes the form (modulo constraints, see below)  \beq
Q= \frac{1}{4m}
\int_{\Sigma} dx (1-2m/R  )^{-1}\sqrt{(dR/dx)^2-q_{xx}(1-2m/R)},\eeq  where
$R
= \sqrt{q_{\theta\theta}}$ and $q_{xx}$ are defined by the line element
  \beq ds^2 = -(N(x,t)\,dt)^2 +
q_{xx}(x,t)(dx+N^x(x,t) \,dt)^2 +q_{\theta \theta}(t,x) (d\theta^2 + \sin^2
\theta d\phi^2)\; .  \eeq The (local) variable $x$ is assumed to coincide
asymptotically ($ x \rightarrow \infty$) with the usual Euclidean radial
variable $r$ (our notation in the present paper is generally the
same as in ref.\ I). \\ For the Schwarzschild solution \beq  q_{\theta\theta}
= x^2, ~~~ q_{xx}= (1-2m/x)^{-1} \eeq we have $Q=0$. Remembering Birkhoff's
theorem on the uniqueness of Schwarzschild's solution in the
context of spherical symmetry\cite{2} one
wonders whether there are "observable" configurations which have $Q\neq 0$.
\\ We
have argued in ref.\ I that such configurations are possible in the
Hamiltonian picture where the diffeomorphisms are to be generated by finite
constraint functionals and cannot be implemented "by hand" as this is done in
the mainly geometrical proof of Birkhoff's theorem  in the conventional
space-time picture. A related observation
 was already discussed by Ashtekar and Samuel\cite{3} for asymptotically
flat Bianchi models. \\The possibility of a nonvanishing observable $Q$ within
the Hamiltonian framework is
associated with the subtleties of slicing the 4-dimensional pseudo-riemannian
manifold into space and time: A nonvanishing  $Q$ requires a nonvanishing
shift $N^x$! \\
This last observation was explicitly discussed in ref.\ I,  but
 we did not try to
give a  concrete physical interpretation of the quantity $Q$. It is the aim of
the present paper to give such an interpretation and discuss some of its
implications. \\ The main result is that the quantity \beq T= 8m Q
\eeq can be interpreted as a time variable canonically conjugate to the mass
$m$! This interpretation does not only follow from the formal canonical
conjugacy of $T$ with respect to the mass $m$ but especially from the
relations \beq dT/dt = \{T,H_{tot}\}_{\bar{\Gamma}} =2N^{(\infty)}(t),~~~~
H_{red} =
H_{tot}|_{ \,\bar{\Gamma}} = 2m\, N^{(\infty)}(t), \eeq  where $H_{tot}$
means the
total Hamiltonian consisting of the (nonvanishing) constraints and the surface
terms (see eq.\ (2.4) below), \{.,.\} denotes the Poisson bracket, $H_{red}$
means
 the value of $H_{tot}$ on the constraint surface
$\bar{\Gamma}$ where the constraints vanish and $N^{(\infty)}(t)$ is the lapse
function $N(t)$ in the spatially asymptotic region under consideration (there
may be several of such regions each one having its own lapse function).
\\  This
means that both Hamiltonians generate time translations as  symmetry
transformations - in contrast to  gauge transformation - in  asymptotically
flat
regions. The dependence of the "observable" $T(t)$ on the "unobservable"
asymptotic time
parameter $t$ is determined by the gauge dependent quantity $N^{(\infty)}(t),
(N^{(\infty)}=1$  for space-times which are asymptotically Minkowski flat). The
factor 2 in eqs.\ (1.5) is due to the normalization of the Hamiltonians adopted
here.  \\ Notice that the observable $T$
is a volume, {\em not} a surface quantity! \\ Our results mean that
spherically symmetric gravity constitutes a completely integrable system with
respect to its observables the mass $m$ being an action, the time $T$ an angle
variable! This resembles the situation in (2+1) gravity\cite{4}. \\ The results
can be translated into the ADM framework where one sees that $T$ is simply
related to an exact Hamilton-Jacobi solution of the (classical)
Wheeler-DeWitt eq.\ for
spherical gravity. \\ Quantization  of the
(1+1)-dimensional canonical system formed by the observables $m$ and $T$ is
straightforward, but depends on the spectrum of $m$, namely whether it covers
the whole real axis or whether it is bounded from below.
\section{The model} We  here
collect the main elements of spherically symmetric gravity in terms of
Ashtekar's variables and refer to ref.\ I for further details. \\
In the spherically symmetric case the  basic canonical variables in Ashtekar's
approach to
quantum gravity, namely the connection coefficients $A^{i}_a(x) $ as
configuration variables and the densitized triads $ \tilde{E}^{a}_i(x),
a=1,2,3, i=1,2,3$ as momentum variables,  can
be expressed by 6 functions $A_I(t,x),  $ and $E^I(t,x), I=1,2,3$, where the
$E^I$ are real and the $A_I$ are complex.   Here $t$
is a "time"
variable and x is a (local) spatial variable which becomes the usual
 Euclidean
radial variable $r$ at spatial infinity. \\ The metric $(q_{ab})$ on the
3-dimensional (Cauchy) surfaces $\Sigma^3$  takes the form \beq
(q_{ab}) =\mbox{diag}(\frac{E}{2E^1}, E^1, E^1 \sin^2\theta ),~ \det(q_{ab})=
\frac{1}{2} E^1 E \sin^2\theta,~ E= (E^2)^2 + (E^3)^2 , \eeq which shows that
the variables $E^1$ and $E$  determine the sign and the degeneracies of the
spatial
metric. \\ Integrating the  Einstein-Ashtekar action over the unit sphere in
$\Sigma^3 = S^2\times \Sigma$, where $\Sigma$ is an appropriate 1-dimensional
manifold (see below), we get the following effective (1+1)-dimensional action
for
spherically symmetric gravity  \begin{eqnarray}  S&=& \int_{R}dt[\int_{\Sigma}
 dx ( -i
E^I \dot{A_I}) -H_{tot}],~ \\ & &\mbox{with the total
Hamiltonian }
 \\ H_{tot}&=&\int_{\Sigma} dx (i\Lambda\,G-iN^x\,(V_x -A_1 G)+ \tiN\, C)+
 Q_r+P_{ADM}
 +E_{ADM},
\end{eqnarray} where \begin{eqnarray} G &=& (E^1)'+
A_2 E^3-A_3 E^2 ~: \\ & & \mbox{Gauss constraint function}; \nonumber \\ V_x
&=&
B^2E^3-B^3E^2
{}~:\\ & & \mbox{vector constraint function}; \nonumber \\ C
&=&(B^2E^2+B^3E^3)E^1+\frac{1}{2}EB^1 ~:\\ & & \mbox{scalar constraint
function};\nonumber \\ (B^1,B^2,B^3) & = &(\frac{1}{2}((A_2)^2+(A_3)^2-2),A_3'
+A_1 A_2,-A_2'+
A_1A_3)~:\\ & & \mbox{"magnetic" fields}\nonumber ; \\ \Lambda &:&
\mbox{Lagrange
multiplier for the Gauss constraint}; \nonumber \\ N^x &:& \mbox{"shift"
(see eq.
(1.2))}
\\ & & \mbox{and Lagrange multiplier for
the vector constraint}\nonumber ; \\ \tiN &=& N/\sqrt{(EE^1/2)}~:~
\mbox{Lagrange multiplier for the} \\ & &\mbox{scalar constraint},~ N~:~
\mbox{lapse
(see eq. (1.2))}; \nonumber \\ Q_r&=&  -i
\int_{\partial \Sigma} \Lambda E^1~: ~ \mbox{charge of
the remaining $O(2)$-symmetry };\\ P_{ADM}&=& \int_{\partial
\Sigma}
i N^x(A_2 E^2+(A_3-\sqrt{2})E^3)~: \\ & & \mbox{ADM-momentum in x-direction};
\nonumber \\
E_{ADM}&=& \int_{\partial
\Sigma} \tiN (A_2
E^3-(A_3-\sqrt{2})E^2)E^1~:~ \mbox{ADM-energy}. \end{eqnarray} As
usual\cite{5,6} the
surface terms arise from the requirement that the 3 constraint functionals
\beq \bar{G}[A,E;\Lambda]= \int_{\Sigma} dx\,\Lambda\, G,~~
\bar{V_x}[A,E;N^x]= \int_{\Sigma} dx\,N^x\, V_x,~~
\bar{C}[A,E;\tiN]= \int_{\Sigma} dx\, \tiN\, C \eeq are functionally
differentiable with respect to $A_I$ and $E^I$.
A dot means $d/dt$ and a prime $d/dx$. \\ The (equal "time") Poisson brackets
are  \begin{equation}
\{A_I(x),E^J(y)\}=i\delta_{I}^{J}\delta(x,y)\;,\;\{A_I(x),A_J(y)\}=\{E^I(x),
E^J(y)\}=0.
\end{equation}  The normalization of the energy $E_{ADM}$ is such that
$E_{ADM}=2m\,N^{(\infty)}, N^{(\infty)} = N(t, x \in \partial \Sigma) $ for
the Schwarzschild mass $m$ (with Newton's constant $G=1,
c=1$). \\ The Hamiltonian $H_{tot}$ generates  gauge transformations
and motions: \beq \delta A_I = \{A_I, H_{tot}\} \epsilon,~~~ \delta E^I =
\{E^I,H_{tot}\} \epsilon ,  \eeq  where $\epsilon$ is a corresponding
infinitesimal parameter.
Explicitly we have
\begin{eqnarray}\delta A_1 & = & [\Lambda'+(N^x A_1)'+i\tiN(B^2 E^2+B^3 E^3)]
\epsilon,
 \\ \delta A_2 & = & [\Lambda A_3+ N^x A_2'+i\tiN(B^2 E^1+B^1 E^2)]\epsilon,
\\ \delta A_3 & = & [-\Lambda A_2+ N^x A_3'+i\tiN(B^3 E^1+B^1 E^3)]\epsilon,
\\ \delta E^1 & = & [ N^x (E^1)'-i\tiN (A_2 E^2+A_3
E^3)E^1]\epsilon,
 \\ \delta E^2 & = & [\Lambda E^3+(N^x E^2)'-i\tiN(A_1
E^1E^2+\frac{1}{2}A_2E)  -i(\tiN E^1 E^3)']\epsilon,
\\ \delta E^3 & = & [-\Lambda E^2+(N^x E^3)'-i\tiN(A_1
E^1E^3+\frac{1}{2} A_3 E)  +i(\tiN E^1 E^2)']\epsilon
\; . \end{eqnarray} Whereas the functions $E^I$ are real, the
connection functions $A_I$ are complex, because \beq A_I= \Gamma_I + i K_I~~
I=1,2,3 ~~ ,\eeq where $\Gamma_I$ are the (reduced) coefficients of the spin
connection and $K_I$ the corresponding coefficients of the extrinsic
curvature. These coefficients can be expressed in terms of the functions
$E^I$:
\beq
(\Gamma_1,\Gamma_2,\Gamma_3)  =  (-\beta',-(E^1)'\frac{E^3}{E},(E^1)'
\frac{E^2}{E}) \;
,~~~\beta'= \frac{E^2 (E^{3 })' -E^3 (E^{2})'}{E} \;,
\eeq where $\beta = \arctan (E^3/E^2)$, and \beq (K_1,K_2,K_3) =
\frac{1}{\tiN E E^1}(E^1(\dot{q}_{xx}-(q_{xx})'N^x-2q_{xx}(N^x)',
  E^2(\dot{E}^1-N^x(E^1)'). \eeq
\section{Topology of the Cauchy surfaces $\Sigma$ and  asymptotic
properties at spatial infinity}
\subsection{Possible topologies}
For spherically symmetric systems the topology of the Cauchy 3-manifold
$\Sigma^3$ is
necessarily of the form $\Sigma^{3}=S^2\times\Sigma$ where $\Sigma$ is a
1-dimensional manifold.\\
 In this paper we are only interested in  topological situations where
 $\Sigma$ is open or compact with boundary,
 especially when it is  asymptotically flat:
As was motivated already in ref.\ I we choose here
\beq \Sigma=\Sigma_n \; ,\; \Sigma_n\cong K\cup\bigcup_{A=1}^n\Sigma_A \; ,
\eeq
i.e.\ the hypersurface is the union of a compact set K (diffeomorphic to a
compact interval) and a collection of ends (each of which is diffeomorphic
to the positive real line without the origin) i.e. asymptotic regions
with outward orientation and all of them are joined to K. This means, we
have n positive real lines, including the origin, but one end of each line
is common to all of them, i.e. these parts are identified. Since the
identity map is smooth, this is still a $C^\infty$ (Hausdorff) manifold.\\
As an example consider the Kruskal-extended Schwarzschild-manifold (see e.g.\
ref.\cite{7}), where we have {\em two} ends $\Sigma_1$ and $\Sigma_2$
each of which
belongs to the asymptotic region $x \to \infty$ with $N_2^{(\infty)}=
-N_1^{(\infty)}=-N^{(\infty)}$. \\
We want to point out here that the boundary of the compactum K has {\em
nothing} to do with the location of
 a horizon, it is just a tool to glue the various ends together and
thus is a kinematically fixed ingredient of the canonical formalism, whereas
the location of a horizon will depend on the mass of the system which is
a dynamical object. In particular, one and the same topological
compactum will be used
for all possible values of the mass. Thus, although it is appropriate to
draw the spacetime pictures which one can find in textbooks for, say, the
Schwarzschild configuration with parameter $m$, the lines $x=2m$ which
seperate the 4 Kruskal regions do not correspond to a specific coordinate
value for the boundary of  the
 compactum K and its time evolution. This property should not give rise to
 confusion
because from the mainly geometrical point of view one is used to the fact
that the
topology of $\Sigma$
 may change under evolution while in the Hamiltonian picture
topology change is excluded by definition. \\ Notice that - according to eqs.
(2.17)-(2.22) - we can have independent evolutions in different ends of
$\Sigma$ by
choosing the support of the Lagrange multipliers $\Lambda, N^x$ and $\tiN$
appropriately. \\
\subsection{Asymptotic properties of the fields} The following discussion of
the asymptotic properties of the fields differs slightly from the one in ref.\
I. It is, however, more appropriate for the physical interpretation of the
observables of the system discussed below. \\
Most important for our purpose are the asymptotically flat manifolds
$\Sigma^3$, for which we have \beq \lim_{r \to \infty} q_{ab} =  q_{ab}^0 +
\frac{f_{ab}(\vec{x}/r,t)}{r} + O(1/r^2), \eeq where $q_{ab}^0$ denotes a fixed
flat Euclidean metric with coordinates $x^a$ and $r^2 = q_{ab}^0 x^a x^b.$ \\
In
our case we have \beq q_{rr} \to 1 + q^r_{-1}/r + q^r_{-2}/r^2 \ldots,~~
q_{\theta
\theta} \to r^2 + q^{\theta}_1 r + q^{\theta}_0 \ldots ~.\eeq We want to
translate this asymptotic behaviour of the metric coefficients into that of
the quantities $E^I$. This leads to the ansatz (compare eq. (2.1))
\begin{eqnarray}
E^1 &\to& x^2 + e^1_1x + e^1_0 + \ldots , \\ E^2 &\to & \sqrt{2} e^2_1 x +
e^2_0 +
e^2_{-1}/x+ \ldots , \\ E^3 & \to &  \sqrt{2} e^3_1 x + e^3_0 + e^3_{-1}/x
\ldots,
(e^2_1)^2 + (e^3_1)^2=1. \end{eqnarray}  These asymptotic relations  have to
be made compatible
with those of the canonically conjugate quantities $A_I$, for which we start
with the ansatz \begin{eqnarray} A_1 & \to & a^1_{-1}/x + a^1_{-2}/x^2 +
\ldots , \\ A_2 &\to& a^2_{-1}/x + a^2_{-2}/x^2 + \ldots, \\ A_3 & \to &
\sqrt{2} +
a^3_{-1}/x + a^3_{-2}/x^2 + \ldots . \end{eqnarray} Because of the eqs.\
(2.23) and (2.24) the asymptotic properties of $E^I$ and $A_I$ are not
independent. They imply the relations (recall that the $E^I$ are real!)
\beq \Re e(a^1_{-1})= 0,~~
 \Re e(a^1_{-2}) = \frac{1}{\sqrt{2}}(e^3_0 e^2_1 - e^2_0 e^3_1),~~
\Re e(a^1_{-3}) = \sqrt{2} e^3_{-1} -e^3_0 e^2_0, \eeq \beq 0=e^3_1~
 (\Rightarrow e^2_1 = 1),~   \Re e(a^2_{-1})  =  -e^3_0,~  \Re e(a^2_{-2})
=  -e^3_{-1} + e^3_0 (\sqrt{2}e^2_0 - \frac{1}{2} e^1_1),\eeq
\beq \Re e(a^3_{-1}) =
\frac{1}{\sqrt{2}} e^1_1 -e^2_0. \eeq  The eqs. (2.23) and (2.25) lead to
identities if we express the time derivatives $\dot{q}_{xx}~ etc.$ by means of
the evolution eqs.\ (2.17)-(2.22). \\ In order to fix the
appropriate asymptotic behaviour of the quantities $E^I$ and $A_I$ we require
\\ i) that - essentially following ref. \cite{6} - the
integrands of the Liouville form
\beq \Theta_L = -i \int_{\Sigma} E^I dA_I \eeq and the symplectic form \beq
\Omega = -i \int_{\Sigma} dE^I \wedge dA_I \eeq behave asymptotically as
$O(1/x^2)$ in order for the
 integrals to converge, and that \\ ii) the Gauss constraint function (2.5)
 vanishes at least as $O(1/x^2)$ at spatial infinity. This is the only use
 we make of
 all the constraints as far as the asymptotic properties are concerned. \\
 The condition $ E^I dA_I \to O(1/x^2) $
implies \beq da^1_{-1}= 0,~  d(a^1_{-2} + \sqrt{2} a^2_{-1})=0,~
e^1_1 da^1_{-2} +da^1_{-3} + \sqrt{2} da^2_{-2} + e^2_0 d^2_{-1}+ e^3_0
da^3_{-1}=0,  \eeq and from $ dE^I\wedge dA_I \to O(1/x^2)$ we get \beq
d(\sqrt{2}
a^1_{-2}+a^2_{-1}) = 0,~~~ d(e^3_0 -\sqrt{2} a^1_{-2}) = 0. \eeq All the
conditions obtained to far can be satisfied by requiring
 \beq a^1_{-1}=0,~~~ a^1_{-2} =0,~~~ a^2_{-1} = 0,~~~ e^3_0 = 0,~~~ a^1_{-3}
 +\sqrt{2}
a^2_{-2} = 0. \eeq It will turn out that these conditions alone do not suffice
to ensure the convergence properties of the quantities we are interested in.
This will, however, be achieved by requiring  the Gauss constraint
function (2.5) to decrease at least as $O(1/x^2)$  at spatial infinity. This
implies the relations
\beq e^2_1  =  1,~~ e^1_1- \sqrt{2}e^2_0  =\sqrt{2} a^3_{-1},~
(\Rightarrow \Im m(a^3_{-1}) = 0,~ \mbox{compare eq. (3.12)}), \eeq  \beq
\sqrt{2}
e^2_{-1} + e^2_0 a^3_{-1}+ \sqrt{2} a^3_{-2} = 0. \eeq The resulting
expansions for $E^I$ and $A_I$ are compatible with the evolution eqs.
(2.17)-(2.22) and the following asymptotic behaviour of the Lagrange
multipiers: \beq \Lambda \to O(1/x^2),~~ N^x \to O(1/x^2),~~
N^{(\infty)} =
O(1), \Rightarrow \tiN \to O(1/x^2). \eeq  This asymptotic behaviour of the
Langrange multipliers means that we allow for a spatially asymptotic $O(2)$
symmetry
and for spatially asymptotic time translations - as opposed to the
corresponding gauge transformations which require a stronger decrease for
$\Lambda$ and $N$ (see ref. \cite{1,6}). In the following we shall not
exploit the asymptotic $O(2)$ symmetry and shall treat it as a gauge
symmetry. {\em However, the possibility of generating time
translations at spatial infinity is of utmost importance for the
interpretation of the theory.}
\\ The asymptotic properties of $E^I$ and $A_I$ yield \begin{eqnarray} E &\to&
2x^2 (1 + \sqrt{2}e^2_0/x +(\frac{1}{2}(e^2_0)^2 +\sqrt{2}e^2_{-1})/x^2
+\ldots ,
\\ A &\to& 2(1+ \sqrt{2} a^3_{-1}/x + (\sqrt{2}a^3_{-2} +\frac{1}{2}
(a^3_{-1})^2)/x^2 +\ldots,\\  B^1 & \to & \sqrt{2}a^3_{-1}/x + O(1/x^2), \\
B^2 &
\to& -a^3_{-1}/x^2 + O(1/x^3), \\ B^3 & \to & (3a^2_{-3} + \sqrt{2} a^1_{-4} +
a^1_{-3}a^3_{-1})/x^4 + O(1/x^5) . \end{eqnarray}  The expansions for $E^1$
and $E$ imply \beq  q_{xx}=\frac{E}{2E^1} \to 1 + (\sqrt{2}e^2_0 -e^1_1)/x +
O(1/x^2). \eeq  Comparing this expression with the asymptotic expansion of the
Schwarzschild solution (1.3) we find \beq \sqrt{2}e^2_0-e^1_1=2m = - \sqrt{2}
a^3_{-1}. \eeq The last equality follows from eq.\ (3.18). \\ It implies for
the ADM-energy (2.13) that \beq E_{ADM} = -\sqrt{2}N^{(\infty)}a^3_{-1} =
2\,m\, N^{(\infty)}. \eeq
For a compact $\Sigma$  it is sufficient to require the fields and Lagrange
multipliers to be
smooth and finite everywhere.
 Obviously, the case of compact topologies is much more easier to
handle from a technical point of view.
\section{Symplectic reduction of the model and its observables}

We first recall some basic facts from the theory of symplectic reduction
(for further details see refs. \cite{8,9}).
It was shown in ref. I  that the present model is a field theory with first
class
constraints. Let $\Gamma,\bar{\Gamma}\;\mbox{and}\;\hat{\Gamma}$ denote
the full phase space, its constraint surface (where the constraints are
 satisfied identically) and its reduced phase space (i.e. the constraint
surface, but points in it are identified provided they are gauge
related). The (local) existence of the latter follows from general theorems
that are valid for first class systems. Let
\beq \iota\; :\; \bar{\Gamma}\rightarrow \Gamma\;\mbox{and}\;\pi\; :\;
\bar{\Gamma}
\rightarrow \hat{\Gamma} \eeq
denote the (local) imbedding  into the large phase space and the projection
onto the reduced phase space respectively. Call the symplectic structures
on the 2 phase spaces $\Omega\;\mbox{and}\;\hat{\Omega}$ respectively. Then
the presymplectic structure on $\bar{\Gamma}$ is defined by the pull-backs
\beq \pi^*\hat{\Omega}:=\bar{\Omega}:=\iota^*\Omega \;. \eeq
 In practice one computes the constraint surface and thus
obtains the imbedding. One then defines the presymplectic structure by the
pull-back under the imbedding. After that one computes the gauge orbits
and obtains the projection. The reduced symplectic structure is then
defined by the pull-back under the projection.\\
On the other hand, if $\Theta\;\mbox{and}\;\hat{\Theta}$ are the symplectic
potentials (Liouville forms) for the symplectic structures, we obtain
\beq d\wedge(\iota^*\Theta-\pi^*\hat{\Theta})=\iota^*\Omega
-\pi^*\hat{\Omega}=0, \eeq
whence $ \iota^*\Theta-\pi^*\hat{\Theta}$ is (locally) exact:
\beq dS:=\iota^*\Theta-\pi^*\hat{\Theta}~. \eeq Here  S is the
Hamilton-Jacobi functional which generates a singular canonical
transformation from the large to the reduced phase space. Replacing
the canonical momenta on $\Gamma$ by the  functional derivatives of S
with respect
to the canonical coordinates on $\Gamma$ solves the constraints
because by doing
so one pulls back the momenta to $\bar{\Gamma}$. Hence, one
way of obtaining the reduced phase space is to solve the Hamilton-Jacobi
equation for constrained systems. This has been done in ref. \cite{9} and also
in ref. I. \\ The relation (4.4) suggests an additional method : it says
that, up to a total differential, one obtains the reduced symplectic
potential simply by inserting the solution of the constraint equations
into the full symplectic potential. \\ For field theories there might also
be boundary terms involved in this reduction process the contribution of
which to
the reduced symplectic structure does not vanish. They may be neglected
at a first stage because they will be recovered when one checks whether
the observables of the reduced phase space are finite and functionally
differentiable.\\
It  turns out that  the last method is quite appropriate for our model. \\
In the following we only discuss the nondegenerate case $ E^1 E \neq 0$.
It then follows that the vanishing of the scalar and vector constraint
functions
$C_x $ and $C$ (eqs. 2.6 and 2.7) is equivalent to the vanishing of the
functions \beq C_2 = B^2 E^1+ \frac{1}{2} B^1 E^2,~~ C_3 = B^3 E^1 +
\frac{1}{2} B^1 E^3, \eeq because \beq E^1 C_x = E^3 C_2 - E^2 C_3,~~ C = E^2
C_2 + E^3 C_3~. \eeq The vanishing of $C_2$ and $C_3$ implies \beq E^2 = g
B^2,~ E^3 = g B^3,~ g= -2 \frac{E^1}{B^1}~. \eeq
Here we exclude the trivial case $B^1 =0$ (see ref.\ I). For our purpose the
following combinations will be of interest, too: \begin{eqnarray} A_3 C_2-
A_2 C_3
&=&
E^1 (B^{1})' + \frac{1}{2}B^1 ((E^{1})' -G)~, \\  A_2 C_2 + A_3 C_3 &=& A
\,\gamma \, E^1 + \frac{1}{2}B^1(A_2 E^2 + A_3 E^3)~, \end{eqnarray} where G
is the Gauss constraint function (2.5) and \beq A \equiv (A_2)^2 + (A_3)^2,~~
\gamma \equiv A_1 + \alpha' = \frac{A_2 B^2 + A_3 B^3}{A}~. \eeq  We here
assume $A \neq 0$, because otherwise  $ B^1 =-1,  (B^{1})' =0$ (eq. (2.8) and
a vanishing expression (4.8) implies  $(E^{1})'=0$
contradicting  eq. (3.4). \\  From eq.\ (4.8) we get the constraint \beq
K_1\equiv \frac{1}{2} B^1 (E^{1})' + (B^{1})' E^1 = 0~, \mbox{ or }
(E^1)'=g(B^1)'~. \eeq  Integration yields
\beq \sqrt{E^1}B^1 = const.~. \eeq From the eqs. (3.4), (3.23) and (3.27) we
obtain \beq \sqrt{E^1} B^1 = -2m~. \eeq  This is our first observable. That
this is so
follows from \beq (\delta (\sqrt{E^1} B^1))|_{\bar{\Gamma}} =
\{(\sqrt{E^1}B^1),
H_{tot}\}|_{\bar{\Gamma}} = 0, \eeq where the eqs. (2.18), (2,19), (2.20),
(4.7) and (4.11) have
been used. \\ For the further discussion it is convenient to make use of the
$O(2)$-symmetry in the (2,3)-"plane" of the variables $(A_2,A_3)$ and
$(E^2,E^3)$ by introducing cylindrical coordinates \beq (A_2,A_3) =
\sqrt{A}(\cos\alpha, \sin\alpha),~~ (E^2,E^3)= \sqrt{E}(\cos\beta,
\sin\beta)~. \eeq  They implies the relations \beq G= (E^{1})' - \sqrt{AE}
\sin(\alpha-\beta) \eeq and \beq E^I dA_I = \pi_{\gamma}
d\gamma +\pi_1 dB^1 + \pi_{\alpha} d\alpha -\frac{d}{dx}(E^1 d\alpha)~, \eeq
where
\beq \pi_{\gamma} = E^1~,~~ \pi_1 = \sqrt{\frac{E}{A}} \cos(\alpha-\beta)~,~~
\pi_{\alpha} = G~ . \eeq  We see that the change of variables is tantamount to
a canonical transformation where one of the new momenta is the Gauss constraint
function $G$! \\ In addition, the r.h.s of  eq. (4.9) now takes the form \beq
A(\pi_{\gamma}\gamma + \frac{1}{2} \pi_1 B^1)~, \eeq implying the constraint
 \beq K_2 \equiv 2
\pi_{\gamma} \gamma + \pi_1 B^1 = 0,~ \mbox{ or } \pi_{1} = g\, \gamma~ .
\eeq Setting $G=0$ we get for the constraint functions $C$ and $V_x$ in terms
of the new variables
\begin{eqnarray} C& =& (1+B^1) \pi_1 K_2 +  \frac{1}{2(1+B^1) }
\pi'_{\gamma} K_1~, \\ V_x &=& -(E^1)'\gamma + (B^1)'\pi_1~. \nonumber
\end{eqnarray}
Dropping  the last term in eq.\ (4.17) which leads to a surface integral to be
ignored (see the remarks following eq.\ (4.4) above) the integrand of the
Liouville form reduced with respect to the Gauss constraint is given by
\beq \pi_{\gamma} d\gamma + \pi_1 dB^1 = - \gamma
d\pi_{\gamma}- B^1 d\pi_1 + d(\ldots)~. \eeq  Using the relation (4.20) we
have \beq \gamma d\pi_{\gamma} + B^1 d\pi_1 = \sqrt{\pi_{\gamma}}\,B^1 d(\frac{
\pi_1}{\sqrt{\pi_{\gamma}}})~, \eeq so that the Liouville form (3.13) finally
reduces to \beq \Theta_L = -i \int_{\Sigma} E^IdA_I = i \int_{\Sigma}
\sqrt{\pi_{\gamma}}\,B^1 d(\frac{
\pi_1}{\sqrt{\pi_{\gamma}}})  + d\hat{S} = m\,dT + d\hat{S}~ , \eeq where
eq.\ (4.13) has
been used and where \beq T= T[\pi_1, \pi_{\gamma}] = \int_{\Sigma} \lambda,~~
\lambda \equiv  -2i \frac{
\pi_1}{\sqrt{\pi_{\gamma}}} ~~.\eeq  The quantity $T$ is our second observable:
\\ From eqs.\ (4.13) and (4.20) we have \beq \lambda = -2i g
\frac{\gamma}{\sqrt{\pi_{\gamma}}} = 4i
\frac{\sqrt{\pi_{\gamma}}}{B^1}\gamma = 8\,m\,\rho ~,~\rho \equiv -i
\frac{\gamma}{(B^1)^2}~~, \eeq and since the asymptotic relations for $A_2,
A_3,A,  B^2, B^3  $ and  $B^1 $ (eqs.\ (3.16), (3.18) and (3.22-3.25)) imply $
\rho \to O(1/x^2)$  the integral $T$ exists. From ref.\ I we know that
$B^1$ is weakly real and $\gamma$ weakly imaginary. So $T$ is weakly real. \\
Most important for our interpretation of $T$ is the relation \beq \delta \rho
=\frac{d}{dx} \left [ N^x \, \rho + \frac{4 m^2 \tiN (B^1)'}{A (B^1)^4}\right ]
\epsilon ~~,  \eeq  which
follows from  applying eqs.\ (2.17)-(2.19) to \[ \delta\gamma = \delta(A_1+
\alpha'),~~
\delta \alpha =(A_2\delta A_3 - A_3 \delta A_2)/A,~~ \delta B^1 = \delta A/2 \]
and using the relations (4.7) and (4.13). The expression in the square
brackets of eq.\, (4.26) approaches the value $N^{(\infty)}/(4m)$ for large
$x$. Combining this with eqs.\ (4.14) and (4.25) we get the important result
\beq \delta T =
\{T,H_{tot}\}\epsilon = 2
\sum_{A=1}^n N_{A}^{(\infty)}\epsilon ~~, \eeq where we have allowed for
different
lapses $ N_{A}^{(\infty)}$ at different ends $\Sigma_A$ (see eq.\ (3.1)) and
where we have assumed that there are no contributions from possible inner
boundary points. The occurrence of the lapses on  the r.h.s.\  of $\delta T$ is
due to our allowance for time translations as a {\em symmetry transformation}
at spatial infinity (\cite{6}), as opposed to a gauge transformation for
which $N^{(\infty)}=0$. \\ If spacetime is  Minkowski-like at spatial infinity
we have $ N^{(\infty)} =1$ and - interpreting the parameter $\epsilon$ as the
(infinitesimal) proper time of an asymptotic observer and considering one
end only- we get \beq \dot{T} = 2, ~~ T = 2t + const. ~~. \eeq Thus, {\em the
observable
T is to be interpreted as a time variable!\/} (The factor 2 is a consequence
of our normalization of the energy - see eq.\ (3.28)). \\ Notice that eq.\
(4.26)
and the eqs.\ (2.17)-(2.22) allow for different - even independent -
evolutions of
$ T $ in
different ends $\Sigma_A $ by choosing the supports of the Lagrangean
multipliers $\Lambda, N^x$ and $\tiN$ (or $N$) appropriately! \\ The mass $m$
is canonically conjugate to the time $T$: Let us define the mass $M =m$  by
\beq M=-\frac{1}{2} \int_{\Sigma} \sqrt{\pi_{\gamma}}\,B^1 \chi,~~
\int_{\Sigma}
\chi =1~~, \eeq where $\chi$ is a suitable smooth test function the support of
which can again be concentrated on a given end $\Sigma_A$. From eqs.\ (3.13)
and (4.17) we  infer the Poisson brackets \beq \{B^1(x), \pi_1(y)\} =
i\delta(x,y),~ \{\gamma(x),\pi_{\gamma}(y)\}= i \delta(x,y),~ \{\pi_1(x),
\pi_{\gamma}(y)\} = 0~ \mbox{etc.} \eeq for the new canonical variables. We
therefore have \begin{eqnarray} \{T,M\}& = & \int_{\Sigma}dx \int_{\Sigma}
dy\, i \{(\frac{
\pi_1}{\sqrt{\pi_{\gamma}}})(x), (\sqrt{\pi_{\gamma}}\,B^1)(y) \chi(y)\}\\ &
=&
 \int_{\Sigma}dx \int_{\Sigma} dy\,
 i\sqrt{\frac{\pi_{\gamma}(y)}{\pi_{\gamma}(x)}} \{\pi_1(x),B^1(y)\}\,\chi(y) =
 \int_{\Sigma}
 dy\, \chi(y) =1~ \nonumber. \end{eqnarray}  {\em We see that we can
 interpret spherical symmetric
 gravity as a (1+1)-dimensional completely integrable canonical system where
 the mass $M$ is the action and the time $T$ the angle variable.}\\ This is in
 complete agreement with the structure of the reduced Hamiltonian $H_{red} =
 H_{tot}|_{\bar{\Gamma}}$ which according to eqs.\ (2.4) and (3.28) takes the
 form \beq H_{red} = 2\,M \, N^{(\infty)}, \eeq because the constraints $G,C_x
$ and $ C$ vanish now and  we have $P_{ADM} = 0$ from our choice of
the asymptotic properties of the fields and of $N^x$. In addition
 we have $Q_r =0$ by
an appropriate choice of the decay ($O( 1/x^{2+\epsilon})$) of $\Lambda$.
Eq.\ (4.31) then yields \beq
\dot{T} = \{T,H_{red}\} =2 N^{(\infty)}~, \eeq in complete agreement with eq.\
(4.27) above. \\ In deriving the results above the introduction of the
variables $\gamma, \pi_{\gamma}$ and $\pi_1$ has been quite useful.
The advantage of these canonical variables will be especially
apparent when solving the Reissner-Nordstr{\o}m model\cite{10}.
\section{Relations to the spacetime metric} We first express $T$ as
functional of the metric coefficients
$q_{rr}$ and $q_{\theta \theta}$ : From eqs.\ (4.16) and (4.18) we get -
without using any constraint - \beq
\pi_{1}^2 = \frac{E}{A} - \frac{((E^1)'-G)^2}{A^2}~. \eeq Assuming $G=0$,
dividing by $E^1$ and
taking the square root yields \beq \lambda  = 2 \frac{1}{A\sqrt{E^1}} (
((E^1)')^2 -A\,E)^{1/2} = 2 (1-2m/R)^{-1}((R')^2 -q_{xx} (1-2m/R))^{1/2}~, \eeq
where the relations \[ q_{xx} = \frac{E}{2E^1}~,~R =\sqrt{q_{\theta \theta}}
 = \sqrt{E^1}~,~ A=2(1+B^1)=2(1-\frac{2m}{R}) \] have been used. The difference
 $ ((E^1)')^2 - EA$ behaves as $O(1/x)$ for large $x$ if the properties (3.21),
 (3.22), (3.18) and (3.19) hold. In that case the integral \begin{eqnarray}
   T[q_{xx},
 q_{\theta \theta}]&=& \int_{\Sigma} 2 (1-2m/R)^{-1} w(q_{xx},
 q_{\theta \theta})~, \\ w(q_{xx},
 q_{\theta \theta}) &=& ((R')^2 -q_{xx}
 (1-2m/R))^{1/2} \nonumber \end{eqnarray} exists because $w \to O(x^{-3/2})$.
  Notice that the decay of the integrand of $T[q_{xx},
 q_{\theta \theta}]$ is determined here without refering to the constraint
 (4.20) which we used in the context of eq.\ (4.25) when calculating the
 asymptotic behaviour of $\lambda$ there! \\ For the quantity $T$ to be real
 the inequality \beq (R')^2 \geq q_{xx}(1-\frac{2m}{R}) \eeq has to be
 satisfied. As was already mentioned above this this is guaranteed due to the
 property of $T$ to be weakly real. \\ For the Schwarzschild solution (1.3)
 $T$ vanishes. How is such  a vanishing $T$ to
 be reconciled with the
 eq.\ (4.27)? The answer lies in the fact that the Schwarzschild-Kruskal
 manifold has $two$ ends for $x \to \infty$ with $N^{(\infty)}_2 =
 -N_1^{(\infty)} $ so that the sum on the r.h.s.\ of eq.\ (4.27) vanishes! It
 is amusing that here a second end is required in order to obtain consistency
 of the formalism! Nonvanishing values of $T$ will be discussed below.\\
 The important eq.(4.27) can be derived in the ADM
 framework, too: It follows from eqs.\ (2.20)-(2.22) that \begin{eqnarray}
 \delta R
 &=& (N^x R'+ \tiN R^2 w)\epsilon \\ \delta q_{xx} &=& [2q_{xx}(N^{x})'
 +q_{xx}'N^x + \frac{1}{w}(2q_{xx}R^2 R'' -q_{xx}'R^2 R'
 -2mq_{xx}^2)\tiN]\epsilon~.
 \end{eqnarray}  In obtaining the last equations the relations \beq \delta R =
 \frac{1}{2R}\delta E^1\,,~~~~~ \delta q_{xx}= \frac{1}{2R^2}\delta E
 -\frac{2q_{xx}}{R}
 \delta R\,,~~\delta E =2(E^2\delta E^2+E^3\delta E^3)\,, \nonumber  \eeq
 have been
 used. For the derivation of eq.\ (5.6) the relations $ E^2(E^3)'-E^3(E^2)'=
 \beta' E$ and \begin{eqnarray} \beta'&=&
 \alpha'-[\arccos\left(\frac{R'}{
 \sqrt{q_{xx}(1-\frac{2m}{R})}}\right)]'~, \\ & & \arccos z
 =-i\cosh^{-1} z = -i\ln(z+\sqrt{z^2-1})~, z\geq 1~,\nonumber \end{eqnarray}
 are essential. The last one follows
 from  eq.\ (4.16) with $G=0$ and observing that $ (\arcsin z)'= (\arccos z)'$
 for $z\geq 1$. \\ The rhs.\ of eq.\ (4.28) arises here from the term \[
 \frac{d}{dx}\left(\frac{\partial \lambda}{\partial R'}\delta R = \frac{2R'}{
w (1-\frac{2m}{R})}\delta R\right)  \] which appears in the integrand when
 $T$ is varied with respect to $R$. Collecting all the terms we get \beq
 \delta T[q_{xx},R]= 2 \int_{\partial \Sigma }(1-\frac{2m}{R})^{-1} (w N^x +
 R^2 R' \tiN)\epsilon ~. \eeq The first surface term vanishes
 and the second one gives eq.\ (4.28). Eq.\ (5.9) is  equivalent  to eq.\
 (4.27) above or eq.\ (5.35) below. \\ The observable $T$ is related in a very
 simple way to the functional \begin{eqnarray} S[q_{xx},R;m) &=&
 \int_{\Sigma}u(q_{xx},R,m) ~, \\ u(q_{xx},R,m) &=& 2R(R'\cosh^{-1}
 \left(\frac{R'}{
  \sqrt{q_{xx}(1-\frac{2m}{R})}}\right) -w)~, \nonumber \\ & & \cosh^{-1}
  \left(\frac{R'}{
  \sqrt{q_{xx}(1-\frac{2m}{R})}}\right) =
 \ln(R'+w)-\frac{1}{2}\ln(q_{xx}(1-\frac{2m}{R}))~, \nonumber \end{eqnarray}
 derived in
 ref.\ I,  which provides an exact Hamilton-Jacobi solution of the (classical)
 Wheeler-DeWitt eq. \beq \frac{1}{R^2\sqrt{q_{xx}}} [ 2
 q_{xx}^2 p_{xx}^2-2q_{xx}Rp_{xx}p_R+2R^2(2R R''+(R')^2-\frac{q_{xx}'R
 R'}{q_{xx}} -q_{xx})]=0 \eeq for spherical symmetric gravity\cite{11}, where
 $p_{xx}$ and $p_R$ are the canonical momenta conjugate to $q_{xx}$ and $R$,
 respectively (the radial function $R$ here should not be confused with the
 curvature scalar $R$ which does not appear in this paper at all!). \\ The
 integral $S$ exists because the integrand $u$ behaves as $ u \to O(x^{-2})$ if
 $w \to O(x^{-3/2})$ for large $x$ as supposed above. Furthermore, since
 \beq \frac{\partial u}{\partial
 R'}=2R[\ln(R'+w)-\frac{1}{2}\ln(q_{xx}(1-\frac{2m}{R}))] \eeq the functional
 $S[R,q_{xx};m)$ is functionally differentiable with respect to $R$  if $
 \delta R \to O(1)$ for $x \to \infty$ because then the surface term \[
\frac{\partial u}{\partial
 R'}\delta R \] vanishes for $x \to \infty$ if $w \to O(x^{-3/2})$.
\\  Inserting  \begin{eqnarray} p_{xx}&=& \frac{\delta S}{\delta
 q_{xx}}= \frac{\partial u}{\partial q_{xx}}= - \frac{w R}{q_{xx}}, \\ p_R &
 =& \frac{\delta S}{\delta R}= \frac{\partial u}{\partial R}-\frac{d}{
dx}\left(\frac{\partial u}{\partial R'}\right)\\ & =& -2w - \frac{1}{w}(2R
 R''-\frac{q_{xx}'R R'}{q_{xx}}-\frac{2m}{R}q_{xx})  \end{eqnarray} into the
 Wheeler-DeWitt eq.\ (5.11) solves that equation identically. In the same way
 the ADM diffeomorphism constraint\cite{11} \beq
 \frac{1}{q_{xx}}(q_{xx}p_{xx}-(2q_{xx}p_{xx})'+R'p_R) =0 \eeq is fulfilled.
 The solution $S(q_{xx},R)$ was also discussed in ref.\cite{12}. \\
 The quantities $S$ and $T$ are related as follows: Eq.\ (4.24) gives
 \beq \Theta_L = -T\,dm + d(\hat{S}-mT)~. \eeq  On the other hand
 we have \beq dS[q_{xx},R;m) = \int_{\Sigma}( \frac{\delta S}{\delta
 q_{xx}}dq_{xx} + \frac{\delta S}{\delta R} dR ) + \frac{\partial S}{\partial
m}
 dm  = \Theta_L + \frac{\partial S}{\partial m} dm~ \eeq Comparing the last
 two equations we can identify $ \hat{S} - mT = S$ and so we have \beq T =
 \frac{\partial
 S}{\partial m}~. \eeq This important relation can be verified by an explicit
 calculation!
  \\ We now come to the discussion of configurations for which the observable
  $T$ does not vanish. Our approach is to start with  given values of the
  observables $T$ and $m$ and
  ask for those values of $N^x$ and $\tiN$ which are compatible with them and
with  the remaining gauge degrees of freedom: \\  We begin with the special
gauge
   $E^1=x^2$ so that $\delta E^1=0\,$! Eq.\ (2.20) implies the condition \beq
 N^x  +\frac{1}{2}x^2 (1-\frac{2m}{x})\lambda \tiN =0 \eeq for $N^x$. It shows
 that $N^x$ has to be nonvanishing for a nonvanishing $\lambda$! \\ The usual
 choice for the Schwarzschild parametrization is $N^x=0$ so that $\lambda =0$
 and $T=0$ in that case.  \\
 We see that a nonvanishing $T$ is associated with a slicing (foliation) of
 spacetime which necessarily has  a nonvanishing lapse $N^x$! \\
 Next we ask what gauge freedom is left in general for
 the gauge dependent functions $N^x$ and $\tiN$ once the observables $m$ and
 $T$ are given: \\ In order to answer this question it is convenient to
 introduce  new canonical variables again by defining \beq s =
 \ln(-B^1)~,~ B^1=-e^{\displaystyle s}~. \eeq  We then have \beq \pi_{\gamma}
 d\gamma +\pi_1
 dB^1 + G d\alpha = i\pi_{\rho}d\rho + \pi_s ds + \pi_{\alpha}d\alpha~,\eeq
 where \beq
 \pi_{\rho}= \pi_{\gamma}(B^1)^2~,~ \pi_{s}= 2\pi_{\gamma}\gamma + \pi_1 B^1~,~
 \pi_{\alpha} = G~, \eeq so that the constraints (see eqs.\ (4.11) and (4.20))
 now take the special form \beq  \pi_{\rho}'=0~,~ \pi_s =0~,~
 \pi_{\alpha}=0~. \eeq  If we define the constraint functional \beq F[\pi_{
 \rho}, \pi_s,
 \pi_{\alpha}] = -\int_{\Sigma} (\lambda_{\rho}\pi_{\rho}' +i \lambda_s \pi_s
 +i \lambda_{\alpha} \pi_{\alpha})~, \eeq  we get from \beq \{\rho(x),
 \pi_{\rho}(y)\} = \delta(x,y)~,~ \{s(x),\pi_s(y)\}=i\delta(x,y)~,~
 \{\alpha(x), G(y)\}=i \delta (x,y) \mbox{ etc.\ }\eeq that \beq \{\rho(x),
 F\}=\lambda_{\rho}'(x)~,~\{s(x),F\}=\lambda_s(x)~,~ \{\alpha(x), F \} =
 \lambda_{\alpha}(x)~. \eeq Thus $F$ generates pure gauge transformations: $F$
 acts additively on the $O(2)$-angle $\alpha$ and the variable $s$ (i.e.\ it
 acts as a scale transformation on $B^1$). Furthermore, if we put $\rho =
 \bar{f}'$ then $F$ acts additively on $\bar{f}$. All the gauge freedom left
 over is now contained in the choice of the variables $\alpha, s$ and
 $\bar{f}$. The choice of $\bar{f}$ is, however, not arbitrary because $ 8m
 \int_{\Sigma} \rho = T$. We therefore introduce \beq \rho = \frac{T}{8m}f'~,~
 \int_{\partial \Sigma} f = 1~,~\mbox{i.e.} \lim_{x\to \infty}f =1~. \eeq This
 is compatible with the property $ \rho \to O(1/x^2)$. \\ Expressed in terms
 of the observables $m$ and $T$ and the  gauge quantities $s$ and $f'$ the
 metric coefficients take the form \begin{eqnarray} q_{xx} &=& \frac{(R')^2}{
 \displaystyle( 1-\frac{2m}{R})}-\frac{1}{4}\lambda^2(1-\frac{2m}{R}) \\ &=&
 4m^2  \frac{
 (s'e{-s})^2}{\displaystyle (1-e^s)} -\frac{1}{4} (Tf')^2 (1-e^{\displaystyle
 s})~,\nonumber \\ q_{\theta \theta} &=&
 4m^2 e^{\displaystyle -2s}~. \end{eqnarray} For the special gauge $q_{\theta
 \theta} =x^2$
 we have $ \exp(-s) = x/(2m) $ and \begin{eqnarray} q_{xx} &=&
 \frac{1}{\displaystyle 1-\frac{2m}{x}} - \frac{1}{4}(Tf')^2
 (1-\frac{2m}{x})~, \\ q_{\theta
 \theta}& =& x^2~. \end{eqnarray} We finally have to express the shift $N^x$
and
 the "lapse" $\tiN$ in terms of $s$ and $f$: Observing that $B^1 =(A/2 -1)$ we
 get from the eqs.\ (2.18) and (2.19) \beq \dot{B}^1 = (B^1)' + m (1+B^1)
 \lambda \tiN~,
 \eeq so that \beq s' N^x + m (1-e^{\displaystyle -s})Tf' \tiN = \dot{s} ~.
 \eeq A second
 equation  for  $N^x$ and $\tiN$ we obtain as follows:
 Eqs.\ (4.14) and (4.26) yield \beq \dot{\lambda} = \frac{d}{dx} \left (
 \lambda N^x + \frac{32 m^3 (B^1)'}{A (B^1)^4} \tiN \right )~. \eeq  Observing
 that $\lambda = T f', \dot{T} =2 N^{(\infty)}$ and integrating
 with respect to $x$ we obtain   \beq Tf'N^x - 16 m^3\frac{s'e^{\displaystyle
 -3s}}{\displaystyle (1-e^s)} \tiN = 2
 N^{(\infty)}f+ T\dot{f}~, \eeq where the "constant" of integration $c(t)=0$,
 due to the asymptotic properties of the different quantities for large $x:
f\to
 1, \dot{f} \to 0, B^1 \to -2m/x , \tiN \to 1/x^2, N^x \to O(1/x^2).$ \\ For
 the special gauge $q_{\theta \theta} = x^2$ we get \begin{eqnarray} x^2\tiN
 &=& \frac{2 N^{(\infty)}f+ T \dot{f}}{2 q_{xx}} \\ \mbox{ or  } N & = &
\frac{2 N^{(\infty)}f+ T \dot{f}}{2 \sqrt{q_{xx}}}~, \\ N^x & = &
-\frac{(1-2m/x)Tf'(2N^{(\infty)}f +T\dot{f})}{2 q_{xx}}~, \end{eqnarray} where
eqs.\ (2.10) and (5.29) have been used.
\section{Quantization} As we have only two physical degrees of freedom, $M$
and $T$, quantization is easy. It depends, however, on the spectrum of the
Mass operator $\hat{M}$, namely whether it is bounded from below or not: \\
1. $ -\infty < m
< + \infty~. $ \\ In this case we have the commutation relations \beq
[\hat{M}, \hat{T}] = -i ~.\eeq  Choosing the representation \beq \hat{T}\phi
(T) = T \phi(T) ~, \hat{M}\phi(T)= -i\frac{d}{dT}\phi(T)~ \eeq with the scalar
product \beq
(\phi_1,\phi_2) = \int_{-\infty}^{+\infty} dT \phi^*_1(T) \phi_2(T) \eeq we
get from eq.\ (4.32) the
 Schroedinger eq.\ for the wave function $\psi(T,t)$:  \beq i\partial_t
 \psi(T,t) =
H \psi(T,t)  = 2N^{(\infty)}\hat{M}\psi(T,t)=
-i2N^{(\infty)}\frac{\partial}{\partial T}
\psi(T,t)~,\eeq  which has the solutions \beq e_m(T,t) = \frac{1}{\sqrt{\pi}}
e^{\displaystyle -im(2\tau(t)-T)} ,~ \mbox{ where } \dot{\tau} =
N^{(\infty)}(t) ~. \eeq However,
in view of the positive energy theorem for isolated gravitational
systems\cite{13} the assumption that the spectrum of $\hat{M}$ is unbounded
from below appears to be unphysical and should be changed. We therefore assume:
 \\ 2. $ 0 < m <
\infty~. $ \\  Here, if the commutation relation (6.1) still holds,
$\hat{T}$ cannot be selfadjoint - and therefore not diagonalizable - because
otherwise $\exp(i\mu \hat{T})$, $\mu$
real, would be a unitary operator which generates translations by $\mu$ in
$m$-space
 violating the spectral condition $m>0$. The problem has been discussed in
 detail by
Klauder et al.\cite{14} and by Isham et al.\cite{15}: \\ If one defines the
operator \beq \hat{S} = \frac{1}{2} (\hat{M}\hat{T}+\hat{T}\hat{M}) ~, \eeq
then we get the Lie algebra commutator \beq [\hat{S}, \hat{M}] = i \hat{M}
\eeq
of the affine group in one dimensions, i.e.\ $\hat{S}$ generates scale
transformations of the spectrum of $\hat{M}$: \beq e^{\displaystyle
-i\beta \hat{S}}\hat{M}e^{\displaystyle i \beta \hat{S}} = e^{\displaystyle
\beta} \hat{M}~, \beta \mbox{ real }~.
\eeq In the space
of functions $\chi(m)$ we may
choose the operator representations \beq \hat{M}=m~, \hat{S}=\frac{i}{2}
(m\frac{d}{dm} +
\frac{d}{dm}m) = i(m\frac{d}{dm}+\frac{1}{2})  \eeq which are selfadjoint with
respect to the scalar product \beq (\chi_1,\chi_2) = \int_0^{+ \infty}dm \,
\chi^*_{1}(m) \chi_2(m) \eeq and where \beq (e^{\displaystyle
i\beta \hat{S}}\chi)(m) =e^{\displaystyle -\frac{1}{2}\beta}
\chi(e^{\displaystyle -\beta }m)~ \eeq leaves
the scalar product invariant. \\ The Schoedinger eq.\ here has the form \beq
i\partial_t \chi(m,t) = 2N^{(\infty)}m \, \chi(m,t)~, \eeq with the solutions
\beq \chi(m,t) = e^{\displaystyle -2im\tau(t)} g(m)~, (g,g) < \infty~. \eeq
The operator
$\hat{S}$ has the eigenfunctions \beq f_{\displaystyle s}(m) = \frac{1}{
\sqrt{2\pi}}
m^{\displaystyle -(i\, s +\frac{1}{2})}~,~ (f_{\displaystyle s_1}, f_{
\displaystyle
s_2}) = \delta(s_1 - s_2) ~, \eeq where $s$ is a real eigenvalue of $\hat{S}$.
\\ A more general choice for the scalar product $(\chi_1,
\chi_2)$ and the operator $\hat{S}$ is \beq (\chi_1,\chi_2)= \int_0^{+\infty}
 dm\, m^{-\sigma}
\chi^*_1(m)\chi_2(m)~, \hat{S}= i(m\,d/dm + (1-\sigma)/2)~, \sigma \mbox{
real}~. \eeq As
to the unitary representations of the affine group in 1 dimension see
ref.\ \cite{16}.
\\ It it easy to verify that the functional $T[\pi_1, \pi_{\gamma}]$ provides
the
solution of Dirac's algebraic quantization approach, too. \\ First we observe
that the constraint functions $K_1(x) $ and $K_2(x)$ in eqs.\ (4.11) and
(4.20) are first class: Let $\chi_i(x), i = 1,2$ be 2 suitable test
functions. Then it follows from the Poisson brackets (4.31) that
\begin{eqnarray}
\{K_1(\chi_1), K_2(\chi_2)\}& =& \int_{\Sigma}dy\int_{\Sigma}dx \chi_1(x)
\chi_2(y) \{K_1(x), K_2(y)\} \\ & =& -i\int_{\Sigma}dx \chi_1(x)\chi_2(x)K_1(x)
=-iK_1(\chi_1\chi_2)~. \nonumber \end{eqnarray} Notice that the constraint
functionals $K_1(\chi_1)$ and $K_2(\chi_2)$ generate the Lie algebra of the
affine group in one dimension, too: $K_2$ generates a scale transformation of
$K_1$. \\  From the Poisson brackets (4.31)
  we infer the (equal time) operator commutation relations \beq
[\hat{\pi}_1(x), \hat{B}^1(y)]=\delta(x,y)~,~ [\hat{\pi}_{\gamma}(x),
\hat{\gamma}(y)]=\delta(x,y)~,~[\hat{G}(x),
\hat{\alpha}(y)]=\delta(x,y)~,~\mbox{ etc. }, \eeq which can be implemented
by the choice
\beq \hat{B}^1= -\frac{\displaystyle \delta}{\displaystyle
\delta \pi_1}~,~ \hat{\gamma}= -\frac{\displaystyle \delta}{\displaystyle
\delta \pi_{\gamma}}~,~ \hat{\alpha}=\alpha~,~   \hat{\pi}_1=
\pi_1~,~ \hat{\pi}_{\gamma} = \pi_{\gamma}~,~ \hat{G}= \frac{\displaystyle
\delta}{\displaystyle
\delta \alpha}~. \eeq Thus the operator versions of the constraint functions
(4.11) and (4.20) take the form \beq \hat{K}_1 = -\frac{1}{2}\pi'_{\gamma}
\frac{\displaystyle \delta}{\displaystyle
\delta \pi_1}- \pi_{\gamma} \left ( \frac{\displaystyle \delta}{
\displaystyle
\delta \pi_1}\right )'~,~ \hat{K}_2 =-2 \pi_{\gamma} \frac{\displaystyle
\delta}{\displaystyle
\delta \pi_{\gamma}}-\pi_1 \frac{\displaystyle \delta}{\displaystyle
\delta \pi_1}~, \eeq where a suitable choice for the operator-ordering in
$\hat{K}_2$ has been made. It is easy to see that $ \hat{G}T=0, \hat{K}_1T=0,
\hat{K}_2T=0$ so that any complex function $\psi(T)$ is annihilated, too. The
first class property (6.16) holds also for the operators $\hat{K}_1$ and
$\hat{K}_2$. \\ \\
Acknowledgements \\
We thank P.\ Hajicek, H.\ Nicolai and H.-J.\ Matschull for stimulating
discussions. One of us (T.T.) is especially indebted to A.\ Ashtekar and L.\
Smolin for many helpful discussions and for their very kind hospitality during
several months at the Physics Department of Syracuse University and at the
Center
for Gravitational Physics and Geometry at Pennsyl\-vania State University. He
is also grateful for a Graduierten-Stipendium of the Deutsche
Forschungsgemeinschaft and for the associated travel funds.

  \end{document}